\documentclass[
    reprint,
    aps
]{revtex4-2}

\usepackage{amsmath}
\usepackage{amssymb}

\usepackage{graphicx}
\usepackage{dcolumn}
\usepackage{bm}
\usepackage{xcolor}
\usepackage[dvipsnames]{xcolor}

\usepackage{makecell} 

\usepackage{array}

\newcolumntype{L}[1]{>{\raggedright\arraybackslash}p{#1}}

\begin{document}

\preprint{APS/123-QED}

\title{Fracture and failure of shear-jammed dense suspensions under impact}

\author{Malcolm Slutzky$^{1,2}$}

\author{Alice Pelosse$^1$}

\author{Michael van der Naald$^{1,2}$}
\author{Heinrich M. Jaeger$^{1,2}$}

\affiliation{
 $^1$James Franck Institute,
 University of Chicago,
 Chicago, Illinois 60637, USA
}

\affiliation{
 $^2$Department of Physics, University of Chicago, Chicago, Illinois 60637, USA
}

\date{\today}

\begin{abstract}

    Impacted with sufficiently large stress, a dense, initially liquid-like suspension can be forced into a solid-like state through a process known as shear jamming.
    While the onset of shear jamming has been investigated extensively, much less is known about the resulting solid-like state in the high stress limit and, in particular, about its failure mode.
    We report on experiments that produce such high-stress failure by impacting dense suspensions with a cylindrical rod moving at controlled speed.
    Using suspensions of cornstarch particles we vary the impact speed over several orders of magnitude and change the fluid viscosity as well as the surface tension in order to identify the conditions for failure.
    The results are compared with similarly dense suspensions of potato starch or polydisperse silica particles.
    In all cases where the shear-jammed suspension fails by fracturing, similar to brittle solids, we observe two types of cracks: a primary circular crack around the impacting rod followed by secondary radial cracks.
    Mapping out the onset of radial fracturing for different particle volume fractions $\phi$ and impact speeds, we identify the requirements for failure via crack formation to occur with at least 50\% likelihood and record the corresponding normal stress $\sigma_\text{N}$ on the impactor.
    We find that this likelihood is not particularly sensitive to changes in particle diameter, but increases when the solvent’s viscosity or its surface tension are reduced.
    In the state diagram for dense suspensions we use these data to delineate the upper limit of shear-jammed rigidity and the crossover into a fracture regime at large $\phi$ and large $\sigma_\text{N}$, several orders of magnitude above the stress required for the onset of shear-jamming.
    We find that the onset of fracturing in many cases is correlated with signatures of internal ductile deformation of the shear-jammed material underneath the impactor, observable in $\sigma_\text{N}$ as a function of axial strain.
    However, for smaller suspension volumes and larger impact speeds, we find strain-hardening up to the point of fracturing.
    This more brittle behavior results in a modulus that, just before crack formation, is roughly an order of magnitude larger than what we observe in shear-jammed suspensions undergoing internal ductile deformation.

\end{abstract}

\maketitle

\section{\label{sec:introduction}Introduction}

    Dense suspensions comprising solid particles dispersed in a liquid solvent at large solid volume fraction display intriguing properties, including a highly nonlinear mechanical response to applied shear \cite{barnes_shearthickening_1989, mewis_colloidal_2011}.
    Above a certain threshold stress, the viscosity of these suspensions can increase by more than an order of magnitude through a process known as discontinuous shear thickening (DST) \cite{barnes_shearthickening_1989, mewis_colloidal_2011, maranzano_effects_2001,fall_shear_2008,brown_role_2012,   wyart_discontinuous_2014, brown_shear_2014, hsiao_rheological_2017, singh_constitutive_2018, morris_lubricated--frictional_2018,ness_physics_2022}.
    At even larger stresses, initially liquid-like suspensions can be forced beyond the DST regime into a solid-like state through shear jamming (SJ) \cite{bi_jamming_2011, peters_direct_2016,  wyart_discontinuous_2014, han_high-speed_2016, han_shear_2018, morris_lubricated--frictional_2018, singh_constitutive_2018,dhar_signature_2020}.
    This behavior of dense suspensions can be mapped in a state diagram as a function of the particle volume fraction $\phi$ and the applied shear stress \cite{ peters_direct_2016, morris_lubricated--frictional_2018, james_interparticle_2018, singh_constitutive_2018, han_stress_2019, hsiao_rheological_2017, wyart_discontinuous_2014}.
    While the transitions into the DST and SJ regimes have been researched extensively in recent years,  less is known about the mechanical properties of the shear-jammed solid at high stress, where the material reaches the limit of its load-bearing capacity.
    Here we report on systematic experiments that investigate how impacting a volume of dense suspension at its free surface can drive the material all the way from its fluid state deep into the SJ regime and eventually produce failure of the dynamically shear-jammed solid. 
    We are particularly interested in the conditions under which this failure leads to the formation of cracks, as in brittle fracture.
    
    Previous studies found mechanical failure of shear-jammed suspensions, in the form of fracture, as an accompaniment to extension \cite{bischoff_white_extensional_2010, andrade_dilatancy_2020, smith_dilatancy_2010, james_interparticle_2018, chen_leveraging_2023, smith_fracture_2015}, gas-driven invasion \cite{ozturk_flow--fracture_2020, lilin_fracture_2024}, as well as low-speed \cite{roche_dynamic_2013, maharjan_constitutive_2018, allen_system-spanning_2018} and high-speed \cite{petel_dynamic_2017} impact.
    Fracture of suspensions can also be generated by slow evaporation, which produces cracks with intermittent growth over long time scales \cite{dufresne_flow_2003,dufresne_dynamics_2006}.
    For low-speed impact under stress-controlled \cite{roche_dynamic_2013} as well as rate-controlled conditions \cite{maharjan_constitutive_2018, allen_system-spanning_2018}, the geometric features of fracture involve a circular ``hole" around the impactor, from which cracks propagate radially outwards over the suspension surface.
    \begin{figure*}
        \centering
        \includegraphics[width=\linewidth]{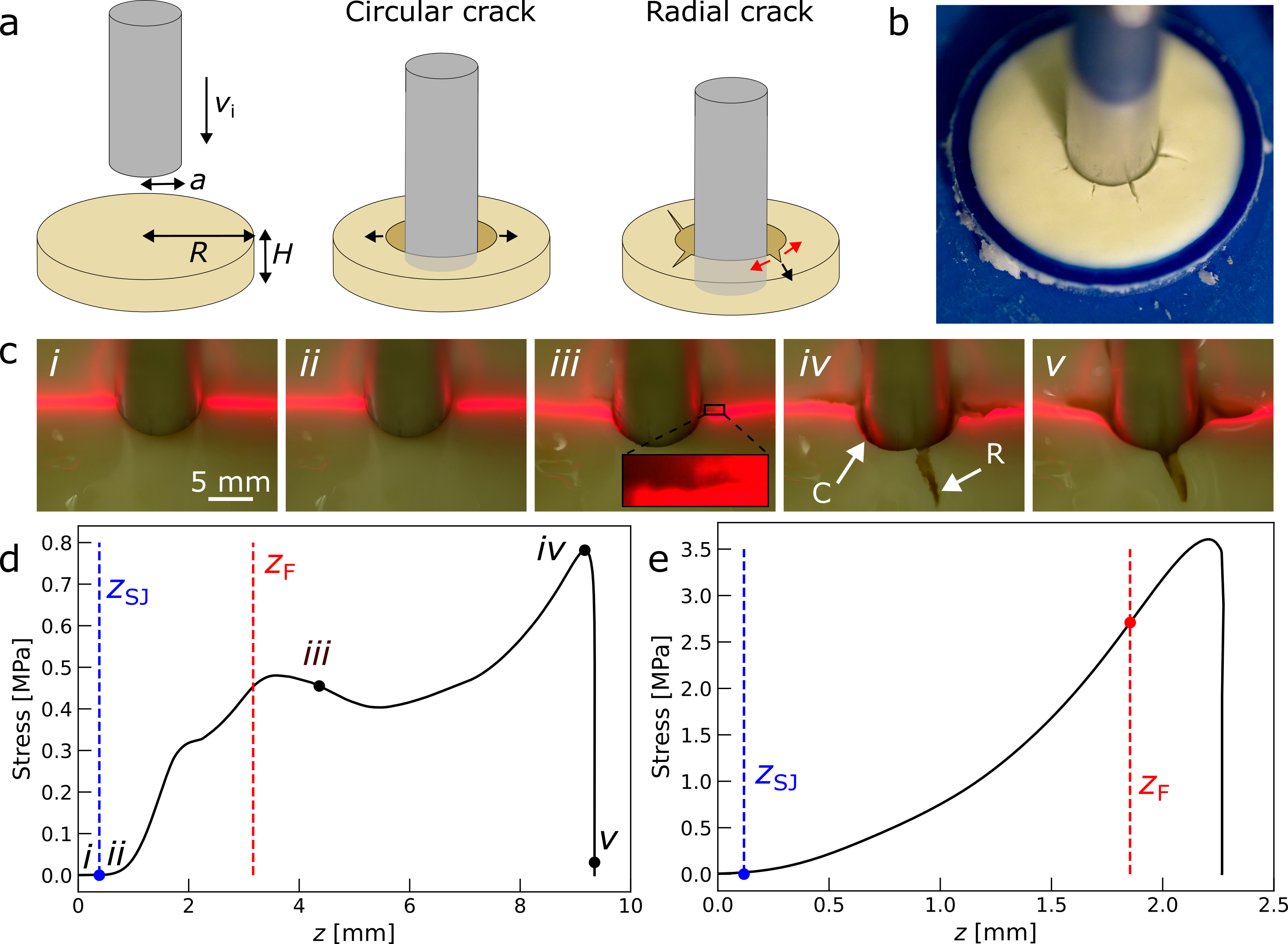}
        \caption{Suspension behavior and typical stress response during impact.
        \textbf{a} Sketches of initial circular fracture and associated hole formation followed by radial cracking.
        Black arrows indicate the widening of the hole and the direction of fracture propagation; red arrows indicate the tension around the hole perimeter.
        \textbf{b} Impact-induced fracturing of a cornstarch suspension in the shallow setup ($v_\text{i} = 10$\,mm/s, $\phi = 0.61$).
        \textbf{c} Still images from a high-speed video tracking a fracture event in the bulk setup for the same suspension and using the same impact speed as in \textbf{a}.
        Inset in \textit{iii}: detail of first radial crack shortly  after it opened.
        Circular (C), and radial (R) fracture are labeled in \textit{iv}.
        \textbf{d} Corresponding normal stress measurement for the bulk setup, with the labels referring to the images in \textbf{c}.
        \textbf{e} Corresponding normal stress measurement for the shallow setup.
        In \textbf{d,e} the depths for shear jamming $z_\text{SJ}$ and fracture $z_\text{F}$ are indicated by blue and red dashed lines.}
        \label{fig:experimental-setup}
    \end{figure*}
    In these cases crack formation is typically akin to mode-I fracture in brittle elastic solids and cohesive granular media \cite{dalbe-crack-2025}.
    When shear-jammed via extension at sufficiently high strain rates, suspensions similarly have been found to fail in a brittle manner \cite{smith_dilatancy_2010,smith_fracture_2015,chen_leveraging_2023, james_interparticle_2018};
    however, extension at lower strain rates typically generates large plastic deformation prior to failure, suggesting a ductile material state \cite{smith_dilatancy_2010,bischoff_white_extensional_2010,smith_fracture_2015,andrade_dilatancy_2020}.
    
    Regardless of how they fail, after the removal of the applied stress, shear-jammed suspensions are able to recover their initial fluid-like behavior\cite{roche_dynamic_2013, smith_dilatancy_2010, maharjan_giant_2017, lilin_fracture_2024}.
    This ability to self-heal after failure together with the frictional, and thus highly dissipative, particle-particle interactions in the SJ regime has led to considerable interest in employing dense suspensions for impact mitigation.
    In particular, fabrics impregnated with a dense suspension in its liquid-like state yield a hybrid material that is soft and flexible in the absence of stress, but transitions to a stiff, solid-like material under impact~\cite{gurgen_shear_2017, sheikhi_16_2023, wang_thickening_2024, lee_ballistic_2003}.
    
    To model the load-bearing capacity of a dense suspension  it is important to consider the coupling of particle movement and flow of the interstitial liquid \cite{jerome_unifying_2016,allen_system-spanning_2018}.
    In any dense granular material consisting of frictionally interacting particles, such as a dense suspension in its shear-jammed state, relative particle movement leads to volume expansion of the particle matrix  (Reynolds dilation \cite{jerome_unifying_2016}.
    In suspensions, this results in a negative pore pressure and causes the interstitial liquid to recede.
    A direct visual consequence is that the surface of a dilated suspension switches from glossy to matte.
    For small amounts of dilation, the negative pore pressure drives the particles further into frictional contact and thus stabilizes the solid-like state.
    However, increased stress loading enhances dilation, and this decreases the number of frictional contacts between a particle and its neighbors, weakening the material and eventually causing it to fail. 
    Where the jammed material is under compression, this failure typically will involve the formation of shear bands, while cracks will open up where tension dominates.
    
    Combining Reynolds dilation with Darcy's Law for the interstitial liquid, Jerome \textit{et al.} showed that it is possible to relate the suspension stiffness, and thus the resistance to penetration, to parameters characterizing the particle matrix, the liquid, and the impact conditions \cite{jerome_unifying_2016}.  
    Although this approach was developed specifically to model suspension behavior near the transition from a fluid-like into a jammed state, it also provides a way to rationalize the very large stresses observed when suspensions are driven deep into the SJ regime.
    These stress levels can reach 1-10MPa with cornstarch   \cite{waitukaitis_impact-activated_2012,allen_system-spanning_2018, maharjan_constitutive_2018} and thus are several orders of magnitude larger than those measured by steady-state rheology near the lower boundary of the SJ regime \cite{han_stress_2019}.
    Still, models capable of predicting the ultimate strength of shear-jammed suspensions are lacking. 
    As a result, no generally applicable criteria are available to predict the failure of dense suspensions at high stress.
    An important step, therefore, is a systematic experimental investigation of the conditions that produce such failure.
    
    In this paper, we address this with experiments in which the suspension is impacted at controlled speed. 
    We focus on cornstarch in aqueous glycerol and adjust variables such as particle volume fraction $\phi$, impact speed $v_{\mathrm{i}}$, liquid viscosity $\eta_0$, and surface tension $\gamma$.
   This is compared with data for potato starch and polydisperse silica spheres.
    To capture the role of confinement and the presence of nearby walls in affecting the strength of the shear-jammed solid, we perform experiments in two different geometries, a bulk setup with a large suspension volume and a shallow setup containing only a relatively thin layer of suspension (slightly thicker than used in typical rheology experiments). 
    
    \textbf{Figure \ref{fig:experimental-setup}} gives an example of an impacted cornstarch suspension.
    Regardless of suspension composition and impact conditions, mechanical failure of the shear-jammed solid proceeds along two consecutive stages (\textbf{Fig. \ref{fig:experimental-setup}a}).
    Initially, the impactor, a rod with flat bottom, creates a circular crack that opens a hole.
    As the hole expands, a hoop strain is generated along its circumference, and this eventually leads to radial cracks extending outward from the rim of the hole.
    \textbf{Figure \ref{fig:experimental-setup}b} shows this hole and the radial fracture in the shallow setup.
    The sequence of images in \textbf{Fig. \ref{fig:experimental-setup}c} shows the radial crack formation in the bulk setup, where we also follow the deformation of the free suspension surface with a laser sheet.

    \begin{table*}[t]
        \centering
        \renewcommand{\arraystretch}{1.5}
        \setlength{\tabcolsep}{8pt}
        \begin{tabular}{l c l c c c}
        \hline\hline
        Particles & \makecell{Particle \\ diameter \\ $d$ [$\mu$m]} & \makecell{Solvent \newline [weight ratio]} & \makecell{Solvent viscosity \\ $\eta_0$ [mPa s]} & \makecell{Surface tension \\ $\gamma$ [mN/m]} & \makecell{Volume fraction \\ $\phi$} \\
        \hline
        Cornstarch (CS1) \textcolor{red}{\textbullet} & $14 \pm 3.6$ & 1:1 glycerol:DI water & 5.1 & 60 & 0.51 -- 0.63 \\
        Cornstarch (CS2) \textcolor{blue}{\textbullet} & $14 \pm 3.6$ & 37:63 CsCl:DI water & 1.0 & 76 & 0.62 \\
        Cornstarch (CS3) \textcolor{Aquamarine}{\textbullet} & $14 \pm 3.6$ & \makecell[l]{1:1 glycerol:DI water \\ + Dawn dish soap} & 5.1 & 45 & 0.61 \\
        Cornstarch (CS4) \textcolor{Mulberry}{\textbullet} & $14 \pm 3.6$ & \makecell[l]{DI water \\ + Dawn dish soap} & 1.0 & 38 & 0.63 \\
        Potato starch (PS) o & $48 \pm 15$ & 1:1 glycerol:DI water & 5.1 & 60 & 0.58 -- 0.60 \\
        Silica spheres (SS) \textcolor{green}{\textbullet} & $5.8 \pm 2.9$ & 1:1 glycerol:DI water & 5.1 & 60 & 0.64 \\
        \hline\hline
        \end{tabular}
        \caption{Particles and solvents used in the impact experiments.}
        \label{tab:suspension-composition}
    \end{table*}

    Tracking the normal (axial) stress on the impactor while it is penetrating the suspension, we record the depths below the initially flat surface and the corresponding stress levels at which we detect shear jamming and the opening up of the first radial cracks (\textbf{Figs. \ref{fig:experimental-setup}d,e}).
    We find that there is a threshold stress well above the onset of shear jamming for suspensions to fracture.
    Below this stress, our impact experiments did not generate radial fracture under any tested condition.
    We also find that there is a minimum volume fraction below which fracture was not achieved.
    Taken together, this allows us to delineate an impact fracture regime.
    Within this regime, our experiments investigate how the probability of achieving fracture, the normal stress on the impactor at fracture, and the hoop strain along the circular hole at fracture depend on impact velocity as well as suspension parameters.  
    We find that impact-induced radial fracturing of the suspension due to increasing hoop strain can proceed regardless of the state of the shear-jammed solid that is being put under compression by the impactor:  
    if this solid exhibits signatures of ductile yielding, as shown in \textbf{Fig. \ref{fig:experimental-setup}d}, radial fracture occurs at the same stress level as ductile yielding.
    However, radial fracture can also occur when, instead of yielding, the compressed solid exhibits continuous strain hardening (\textbf{Fig. \ref{fig:experimental-setup}e}).
    The minimum stress for observing radial cracking in this case is higher, and this correlates with a larger effective modulus of the solidified suspension.
    We discuss how these results put constraints on various scenarios for stress-induced failure in dense suspensions.
    Taking the onset of fracture as indicator for the strength of the shear-jammed material, our data provide an upper boundary delimiting the SJ regime at high packing fraction.
    Finally, augmenting this with literature data for different means of inducing failure, we outline a general fracture regime as a function of stress and packing fraction.

\section{\label{sec:Experimental Procedure}Experimental Procedure}

\begin{figure*}
        \centering
        \includegraphics[width=1\linewidth]{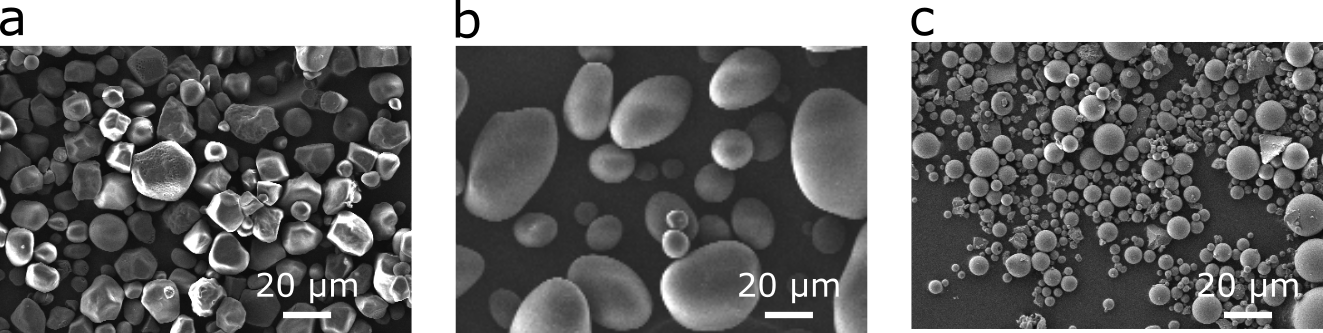}
        \caption{Scanning electron microscope images of the particles used.
        \textbf{a} Cornstarch, \textbf{b} potato starch, and \textbf{c} silica spheres.}
        \label{fig:sem}
    \end{figure*}

    The dense suspensions tested consisted of cornstarch (Sigma Aldrich, S4126), potato starch (Sigma Aldrich, S4251), or silica spheres (Sigma Aldrich, 440345) suspended in solvents composed of DI water mixed with glycerol (McMaster-Carr), dish soap (Dawn, 0.53 g per 100 g of DI water or aqueous glycerol solution), and/or CsCl (Sigma Aldrich 746487).
    The formulations of each type of suspension used are provided in \textbf{Table \ref{tab:suspension-composition}}.
    Particle diameters were measured using a scanning electron microscope (Carl Zeiss Merlin) with accelerating voltage 5.0 kV.
    Representative images of the particles in their dry state are provided in  \textbf{Fig. \ref{fig:sem}}.

    Unless otherwise specified, experiments were conducted with cornstarch suspended in a mixture of glycerol and DI water at a 1:1 weight ratio (labeled CS1).
    The resulting suspensions had particle volume fraction $0.51 < \phi < 0.64$.
    Other types of suspensions were used to test the effect of changing specific suspension properties, such as solvent viscosity (CS2, CS4), surface tension (CS2, CS3, CS4), particle size (PS, SS), or particle type (PS, SS).
    
    To prepare each sample, all the components were weighed and mixed until fully combined.
    To account for the porosity of the starch particles as well as their initial moisture content, the conversion from particle weight fraction to volume fraction $\phi$ developed by Han \textit{et al.} \cite{han_measuring_2017} was used (see Appendix A).
    We employed suspensions with a range of volume fractions, accounting for the differences in volumetric jamming packing fraction $\phi_\text{J}$ between different types of suspensions (see Appendix B).
    Between all suspensions in which we observed fracture, the ratio $\phi / \phi_\text{J}$ did not vary by more than 6\%.
    At the time of preparation, the suspensions were not degassed; the possible left-over bubbles were removed via light mixing with a spoon prior to each experiment. 
    To prevent evaporation, the suspensions were stored in sealed containers, and all experiments were conducted within 2 - 48 hours of a suspension's preparation.
    Over this period of time, the rheological properties of a sealed suspension did not change.
        
\begin{figure}[h]
        \centering
        \includegraphics[width=1\linewidth]{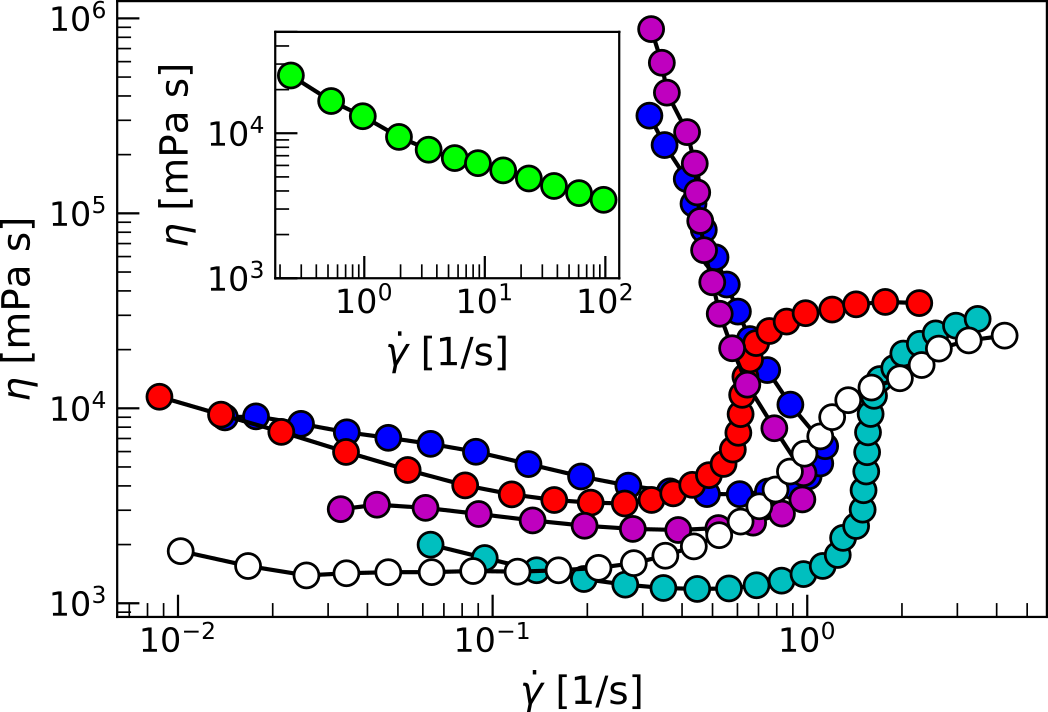} \caption{Suspension rheology.
        Suspension viscosity $\eta$ as a function of shear rate for cornstarch CS1 with $\phi = 0.61$ (red $\textcolor{red}{\bullet}$), CS2 with $\phi = 0.62$ (dark blue $\textcolor{blue}{\bullet}$), CS3 with $\phi = 0.61$ (light blue $\textcolor{Aquamarine}{\bullet}$), CS4 (purple $\textcolor{Mulberry}{\bullet}$) with $\phi = 0.63$, potato starch PS (white o) with $\phi = 0.60$, and silica spheres SS with $\phi = 0.64$ (green, inset $\textcolor{green}{\bullet}$), see \textbf{Table \ref{tab:suspension-composition}}.}
        \label{fig:rheology}
    \end{figure}

    Conducting steady-state stress sweeps with a rheometer (Anton Paar, MCR 301), we confirmed that the starch suspensions were exhibiting strong shear thickening (\textbf{Fig. \ref{fig:rheology}}).
    The differences between the cornstarch suspensions (curves CS1-4 in \textbf{Fig. \ref{fig:rheology}}) may come from different particle surface interactions or (mild) particle swelling depending on the solvent.
    Suspensions made with polydisperse silica spheres required a non-zero shear rate to yield, above which they exhibited mild shear-thinning behavior, see \textbf{Fig. \ref{fig:rheology}} inset.
    Shear-thickening of such suspensions was not measurable within the stress range accessible with the rheometer, but could occur at the higher stresses applied during the impact experiments.

    For impact experiments, a suspension was poured into one of two containers, designated as the shallow setup and the bulk setup.
    Both containers were printed in Vero material using a 3D printer (Stratasys J850).
    Dimensions of the two setups, as marked in \textbf{Fig. \ref{fig:experimental-setup}a}, are provided in \textbf{Table \ref{tab:compare-setups}}.
    The suspension was then impacted from above by a cylindrical, flat-bottomed aluminum rod of radius $a = 5.0$ mm.
    To achieve speed-controlled impacts, the cylindrical impactor was attached to a Materials Testing Machine (ZwickRoell EZ001) with a 1 kN load cell.
    Some data using the bulk setup were obtained using a linear actuator (Dyadic SCN5) with a 50N or 500N load cell (Omega DLC101).
    The impact speeds could be adjusted over 2-4 orders of magnitude, depending on the setup (see \textbf{Table \ref{tab:compare-setups}}).

    \begin{table}[h]
        \def\arraystretch{1.5}
        \begin{tabular}{ccccc}
            \hline
             Container& \hspace{1mm}$R$ [mm] \hspace{1mm} & $H$ [mm] & \hspace{1mm}$Z$ [mm]\hspace{1mm} & \hspace{1mm}$v_\text{i}$ [mm/s] \hspace{1mm} \\
            \hline
            \hline
            Bulk & 45 & 15 & 9.0 - 10 & 1.0 - 100 \\
            Shallow & 15 & 3.0 & 2.0 - 2.5 & 0.001 - 10 \\
            \hline
        \end{tabular}
        \caption{Dimensions and impact conditions for the bulk and shallow setups, with container radius $R$, suspension fill height $H$, final depth of impactor $Z$, and impact velocity $v_\text{i}$.}
        \label{tab:compare-setups}
    \end{table}
    The two  setups were utilized to probe different aspects of the shear-jammed suspensions' properties at fracture, specifically ductile vs. brittle failure modes, as discussed in the next sections.
    The shallow setup provided advantages in terms of a quick turn-around between trials with freshly poured samples.
    Samples in the shallow setup were also more easily mixed with a spoon to remove bubbles.

    In both setups, final impact depth $Z$ and impact velocity $v_\text{i}$ were limited by the instrumentation, as impacts with large $Z$ or $v_\text{i}$  could yield forces approaching the capacity of the load cell.
    In the shallow setup, the accessible range of speed-controlled impacts at fixed $v_\text{i}$ was further limited by the distance required to decelerate the impactor.
    For an impact with $v_\text{i} = 10$ mm/s, all deceleration took place in the last 0.3 mm of the impact, meaning that for $Z = 2.0 - 2.5$ mm, at least $85$\% of the impact was executed at the intended speed.
    
    The experiments were illuminated from above and imaged with a camera (Sony ILCE-1, $3840 \times 2160$ pixels, 120 fps, 96 dpi) mounted $60^\circ$ above the horizontal.
    To highlight the deformation of the suspension surface, a laser sheet (Exor Robotics, SYD1230) was mounted directly across from the camera and focused on the center of the suspension in select experiments.
    The deformation of the suspension surface was calculated by analyzing the displacement of the laser sheet, correcting for distortion due to the $60^\circ$ camera angle.
    This enabled us to detect changes in the local surface elevation with 0.25 mm resolution.

\section{\label{sec:Results}Results}

    \textit{Visual Characterization} - \textbf{Figure \ref{fig:experimental-setup}} shows images of fracture produced in the same suspension (CS1, $\phi = 0.61$)? impacted at the same velocity (10 mm/s) in the shallow setup (\textbf{b}) and the bulk setup (\textbf{c}).
    In both setups, we observe circular and radial fracture, with radial cracks distributed roughly evenly around the impactor.
    
    The sequence of snapshots in \textbf{Fig. \ref{fig:experimental-setup}c}, taken from a video of the bulk setup, shows in more detail how the impact transforms the suspension from fluid- to solid-like and initiates fracture.
    In image (\textbf{c.\textit{i}}), the impactor is about to make contact with the suspension ($z = 0$ mm, $t = 0$ s) and the surface of the suspension is still uniformly glossy, indicative of its fluid state.
    Image (\textbf{c.\textit{ii}}) shows the suspension as the impactor has penetrated to depth $z_\text{SJ}$ (see below for the definition of $z_\text{SJ}$ from the stress measurement).
    Up to this stage, the fluid surface has been deforming slightly under the impactor, as seen from the curvature in the laser line.
    Now the suspension has transformed locally into its solid-like state via rapidly propagating shear-jamming fronts \cite{waitukaitis_impact-activated_2012,peters_direct_2016,han_high-speed_2016,allen_system-spanning_2018}, and  a gap is opening between the suspension and the impactor (circular fracture).
    The suspension surface close to the impactor has turned matte, indicative of the dilation that accompanies shear jamming \cite{waitukaitis_impact-activated_2012, roche_dynamic_2013, han_shear_2018, brown_shear_2014}.
    As the impactor penetrates further and the hole widens, the first radial cracks  along the hole perimeter appear at depth $z_\text{F}$. 
    Image (\textbf{c.\textit{iii}}) shows the suspension shortly thereafter ($z \approx 4.3$ mm, $t \approx 0.43$ s).
    At the time of image (\textbf{c.\textit{iv}})? the impactor experiences its maximum stress during the experiment ($z \approx 9.1$ mm, $t \approx 0.91$ s), before it is decelerated to a stop at its final depth $Z$.
    Radial cracks have propagated outward from the impactor to their maximum length and are long and wide, with plastic deformation in their vicinity.
    The gap between the sides of the impactor and the surrounding suspension has widened significantly.
    In image (\textbf{c.\textit{v}}), the impactor has been left to sit at its final depth ($Z \approx 9.2$ mm, $t \approx 25$~s) while the suspension relaxes.
    At this stage the radial cracks have decreased in length, and their tips have become blunted.
    Regions of the suspension which solidified during impact have begun to return to a liquid-like state, filling in the radial cracks, and
    the portion of the suspension surface local to the impactor and to the cracks has regained its glossy texture.

    The fracture process in the shallow setup proceeds in a very similar manner.
    However, the radial cracks are often more numerous, more jagged, thinner, and shorter than in the bulk setup.
    Unless the impact speed is very low, the entire suspension surface turns matte, indicating that most, if not all, of the suspension volume is in its shear-jammed state.

    \textit{Normal Stress} - Example traces of normal stress on the impactor as a function of penetration depth are shown in \textbf{Fig. \ref{fig:experimental-setup}d,e}  for the bulk and shallow setups, respectively.
    Focusing on the initial regime for both graphs, the background stress is slightly positive as the liquid suspension shear thickens under impact.
    Beyond a characteristic depth $z_\text{SJ}$ (blue dashed lines in \textbf{Fig. \ref{fig:experimental-setup}d,e}) the stress then increases significantly as an accompaniment to shear jamming, signaling the solidification of the dense suspension.
    We iteratively solve for the depth $z_\text{SJ}$ at which the stress rises above the background stress following a procedure introduced by Maharjan \textit{et al.} \cite{maharjan_constitutive_2018}, which takes into account the portion of the impact to which an added mass model is applicable, using the width of the impactor, the depth of the container, and the density of the suspension.
    In our experiments $z_\text{SJ}$ ranged between $0.20 - 3.0$ mm in the bulk setup and $0.10 - 0.90$ mm in the shallow setup.
    
    Once in the shear-jammed state,  further penetration puts the material underneath the impactor under increasing compressive stress and eventually radial crack formation is initiated.
    In our experiments, this radial crack formation occurs at a depth $z_\text{F}$ (red dashed lines in \textbf{Fig. \ref{fig:experimental-setup}d,e}).
    In some experiments, such as the one shown in \textbf{Fig. \ref{fig:experimental-setup}d}, the onset of radial fracture occurs just before the normal stress exhibits a dip,  which is then followed by strain stiffening before the impactor reaches its final depth $Z$.
    This behavior, involving either a slight stress dip or an extended stress plateau, indicates material weakening due to significant plastic deformation within the material compressed underneath the impactor and appears in experiments in the bulk setup as well as for low-velocity impacts in the shallow setup.
    In contrast, normal stress curves for larger impact speeds in the shallow setup, such as in \textbf{Fig. \ref{fig:experimental-setup}e}, show no sign of internal yielding, and instead, increase monotonically, exhibiting continuous strain stiffening up to the largest stress levels applied in our experiments. 
    As the impactor penetrates into the material, there still is significant dilation around the hole and cracks open up at $z = z_\text{F}$.
    In these cases, the normal stresses at the onset of fracturing are significantly higher (by a factor of approximately 5).
     
\begin{figure}[t!]
        \centering
        \includegraphics[width=\linewidth]{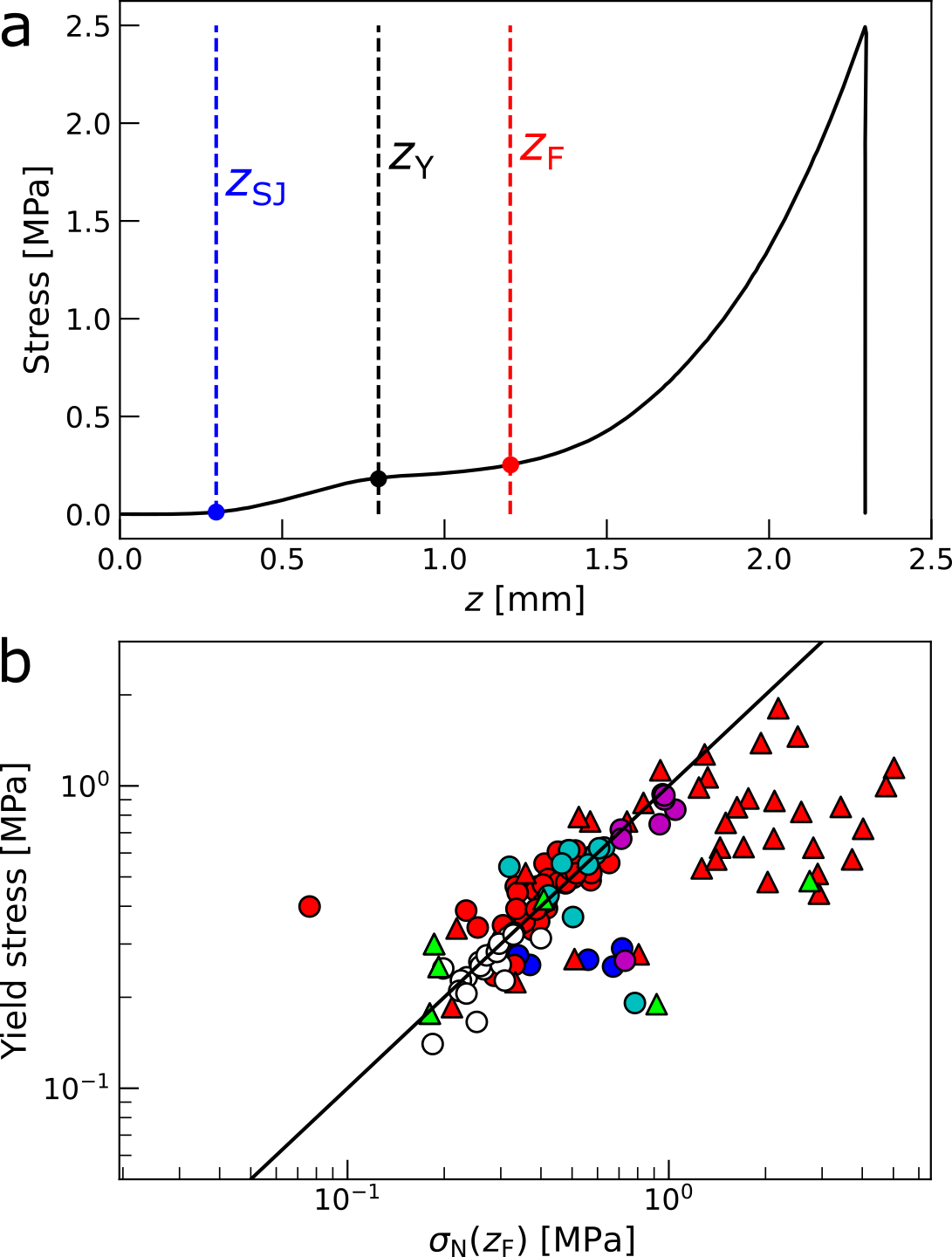} \caption{\textbf{a} Normal stress $\sigma_\text{N}$ versus axial penetration depth $z$ in the shallow setup at low impact speed ($v_{\mathrm{i}} = 0.1$ mm/s) showing internal yielding at $z=z_\text{Y}$ prior to crack formation at $z=z_\text{F}$.
        \textbf{b} Correlation between yielding stress and fracture stress.
        Data are from all trials where yielding was detected in the stress-strain curves for cornstarch CS1 (red $\textcolor{red}{\bullet}$), CS2 (dark blue $\textcolor{blue}{\bullet}$), CS3 (light blue $\textcolor{Aquamarine}{\bullet}$) and CS4 (purple $\textcolor{Mulberry}{\bullet}$), as well as potato starch PS (white o) and silica SS (green $\textcolor{green}{\bullet}$) in the shallow (triangles) and bulk (circles) setups.}
        \label{fig:yielding}
    \end{figure}

     If the impact speed is reduced in the shallow setup, we find a transition to internal yielding behavior, seen by the appearance of a plateau or decrease in slope in the stress-strain curves at a depth $z = z_\text{Y}$, and a concomitant decrease in the normal stress at fracture such as in \textbf{Fig. \ref{fig:yielding}a}.
     Wherever we can clearly identify such internal yielding in stress data, i.e., a stress dip or a stress plateau, we find that the corresponding yield stress correlates well with the fracture stress, generally occurring at slightly higher or lower stresses (\textbf{Fig. \ref{fig:yielding}b}).
     For all bulk cornstarch data, the yielding and fracture stresses have, on average, the same magnitude (black line), while the fracture stress appears to be enhanced relative to the yield stress for much of the shallow cornstarch data.
     Therefore, when we see fracture on the surface of the suspension, there is in most cases a simultaneous signature in the stress curve indicating yielding beneath the impactor.

    \begin{figure}
        \centering        \includegraphics[width=0.8\linewidth]{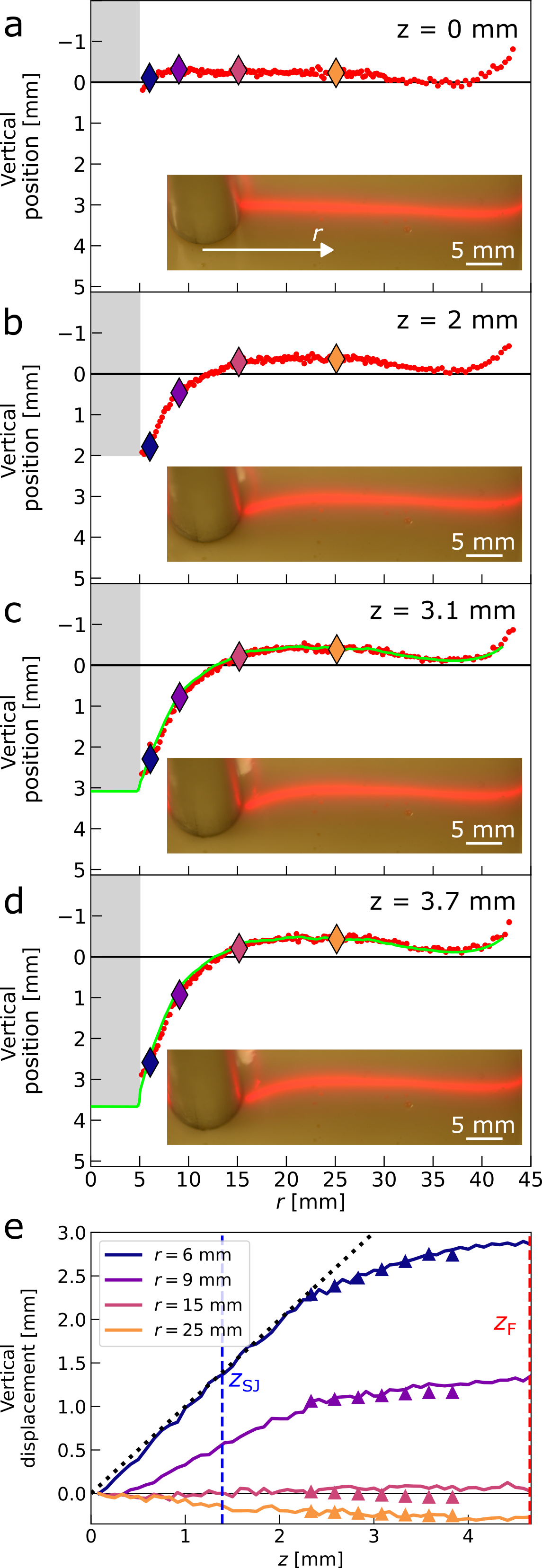}
    \end{figure}

        \begin{figure}
        \centering
        \caption{Suspension surface profile during impact.
        \textbf{a-d} Surface elevation as a function of radial distance $r$ from the axis of the impactor for a CS1 suspension in the bulk setup ($v_\text{i} = 10$ mm/s) at different penetration depths $z$. Insets: images of the suspension surface with the laser sheet used for profile reconstruction. Gray shaded area: position of the impactor. At $z = 0$ the impactor has made first contact with the suspension.
        Red data: surface elevation, green lines: COMSOL simulation after shear jamming has commenced (panels \textbf{c,d}).
        \textbf{e} Vertical displacement from the initial, quiescent surface at four radial positions marked by the diamonds in \textbf{a-d}.
        Triangles: results of COMSOL simulation with bulk modulus 1 GPa and shear modulus 1 MPa.
        Vertical dashed lines: shear-jamming depth $z_\text{SJ}$ (blue) and fracture depth $z_\text{F}$ (red).
        Black dotted line: depth of the impactor.}
        \label{fig:laser-fig}
    \end{figure}

    \textit{Surface Deformation} - 
    In the shallow setup at impact speeds of 1 mm/s or more, the rapid spreading of the shear-jamming fronts quickly leads to complete solidification of the suspension volume.
    This is seen not only by the free surface turning matte across its full extent, but by minimal surface deformation (except for the hole punched by the impactor).
    Analysis using the laser sheet showed that no point of the surface profile in the shallow setup was displaced by more than 0.75 $\pm 0.25$ mm from its initial position.
    
    In the bulk setup, by contrast, the state transition of the suspension under impact was  easily detectable by tracking how the free surface deformed around the impactor.
    Profiles of a bulk CS1 suspension surface at four different impactor depths are plotted in \textbf{Fig. \ref{fig:laser-fig}a-d}.
    The position of the impactor is indicated by the shaded gray rectangle.
    There is a meniscus between the suspension and the container wall, which affects the suspension's profile for radial distances $r > 35$ mm.

    As the impactor moves into the bulk suspension, the material underneath and around the impacting rod is forced downward.
    Initially, for $z<z_\text{SJ}$, the liquid-like suspension surface next to the rod is pinned to the impactor and moves with its bottom face (\textbf{Fig. \ref{fig:laser-fig}a,b}).
    Around $z = z_\text{SJ}$, the suspension locally undergoes shear jamming, its surface depins, and the impactor punctures the underlying material without further distorting its surface (\textbf{Fig. \ref{fig:laser-fig}c,d}).
    
\begin{figure}[t!]
        \centering
        \includegraphics[width=\linewidth]{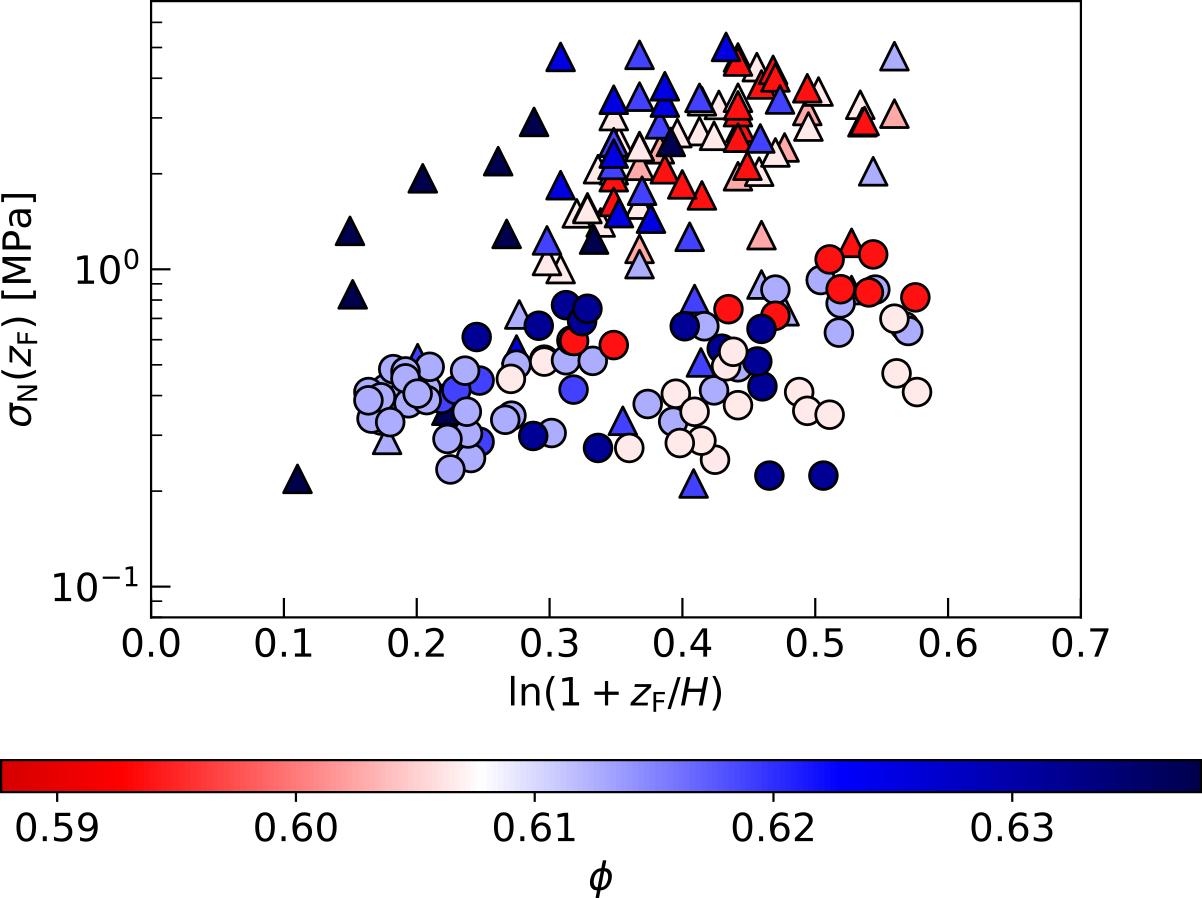} \caption{Normal stress $\sigma_\text{N}(z_\text{F})$ and true axial strain $\ln(1+z_\text{F}/H)$ at the onset of radial crack formation.
        Data are from all trials with cornstarch CS1 in the shallow (triangles) and bulk (circles) setups.
        Volume fraction $\phi$ is indicated by the color of the markers.}
        \label{fig:hist-and-scatter}
    \end{figure}

    To highlight this, we plot the vertical displacement of the suspension surface (relative to the profile before impact) at several radial positions in \textbf{Fig. \ref{fig:laser-fig}e}.
    These positions are indicated with colored diamonds in panels \textbf{a-d}.
    The vertical dashed blue line marks the depth $z_\text{SJ}$, and the vertical red dashed line at the right limit of the plot is the depth at which fracture occurs $z_\text{F}$.
    The dotted black line represents the impactor depth, and the zero vertical displacement is indicated by the horizontal black line.
    The vertical displacement of the suspension at radial position $r = 6$ mm (1 mm away from the impactor) follows the path of the impactor during the initial portion of the impact, as seen by how closely the dark blue data trace overlaps with the black dotted line for $z < z_\text{SJ}$.
    At larger depths, $z>z_\text{SJ}$, the slope of the blue data trace decreases, and the vertical displacement plateaus.
    This trend, of steadily increasing vertical displacement followed by a plateau, applies for $r < 15$ mm (purple data trace).
    For $r = 15$ mm, there is no vertical displacement of the surface during impact (pink trace).
    Apart from the meniscus, the portion of the surface with $r > 15$ mm (orange trace) experiences negative displacement, i.e., in this convention an upward motion, which is due to conservation of suspension volume.
    In the regime of puncture behavior (i.e., when the displacement of the suspension close to the impactor deviates from the impactor's motion), we model the impact using a COMSOL simulation, approximating the jammed suspension as a Neo-Hookean solid.
    The results of this simulation are plotted with triangles and agree with the experimental data for each of the radial locations shown in \textbf{Fig. \ref{fig:hist-and-scatter}e}.

    \textit{Onset of Radial Fracture} - The scatterplot in \textbf{Fig. \ref{fig:hist-and-scatter}} displays the normal stress at fracture onset $\sigma_\text{N}(z_\text{F})$ as a function of axial strain.  
    Given the large relative penetration depths $z_\text{F}/H$, we plot here the true axial strain $\ln(1+z_\text{F}/H)$.  Data shown are for all trials using cornstarch CS1 suspensions in the shallow (triangles) and bulk (circles) setups.
    Volume fraction $\phi$ is indicated by the color of the markers.
    Experiments with volume fraction $\phi < 0.59$ did not yield fracture under any conditions tested.
    
\begin{figure}[h]
        \centering
        \includegraphics[width=\linewidth]{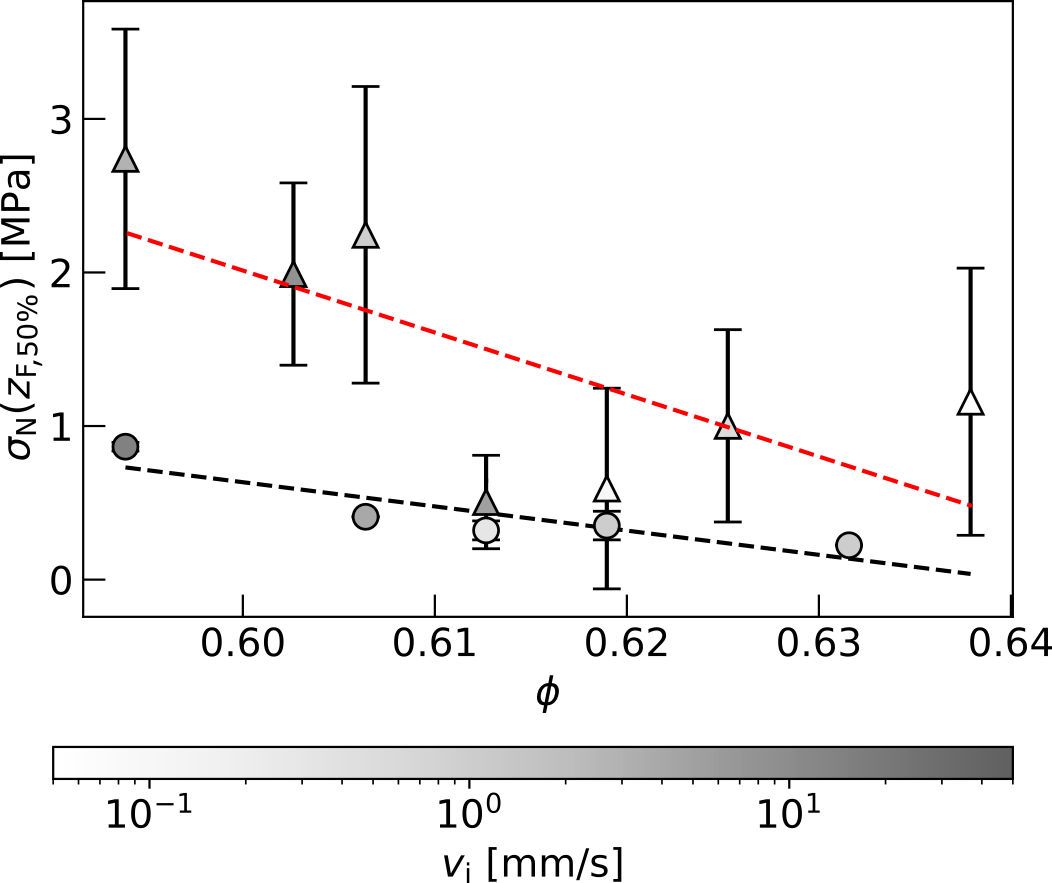}
        \caption{Onset stress for fracture as a function of volume fraction.
        The plot shows the stress $\sigma_\text{N}(z_\text{F,50\%})$ at which fracture occurs with 50\% likelihood, as controlled by the impact speed $v_\text{i}$ (indicated by color) in the shallow (triangles) and bulk (circles) setups.
        The dashed black and red lines are linear fits to $\sigma_\text{N}(z_\text{F,50\%})$ for the bulk and shallow setups, respectively.}
        \label{fig:min_stress_at_fracture}
    \end{figure}

\begin{figure}[h]
        \centering
        \includegraphics[width=0.7\linewidth]{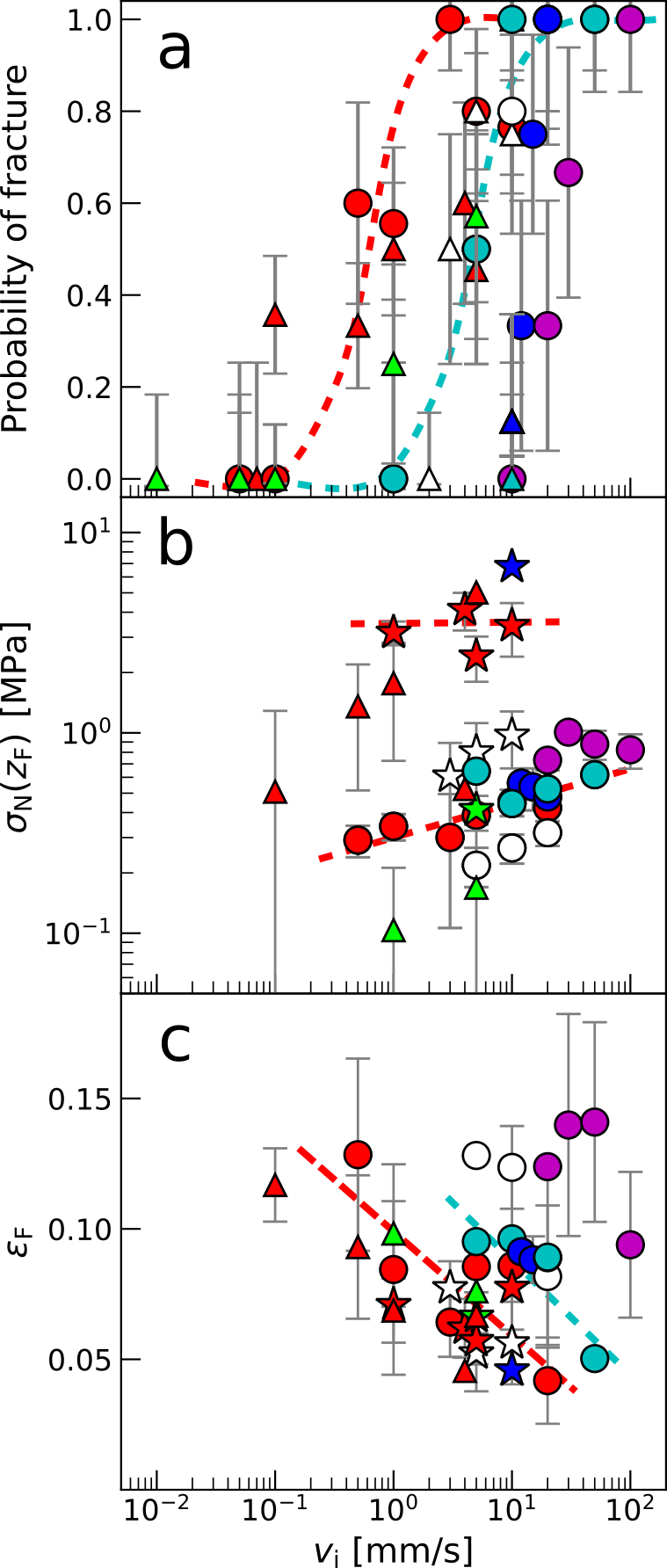}
        \caption{Impact speed dependence of $\phi$-averaged quantities for radial fracture in different suspensions. \textbf{a} Probability of radial crack formation. \textbf{b} Normal stress at radial crack formation $\sigma_\text{N}(z_\text{F})$.  Stars indicate stress-strain curves that only showed strain-stiffening and no signs of ductile failure.
        \textbf{c} Hoop strain $\varepsilon_\text{F}$ at fracture as a function of impact velocity $v_\text{i}$.
        The type of setup is indicated by the shape of the markers, with shallow data in triangles and bulk data in circles.
        Data symbols are for cornstarch CS1 (red $\textcolor{red}{\bullet}$, averaged over 0.61 $<\phi<$ 0.62), CS2 (dark blue $\textcolor{blue}{\bullet}$), CS3 (light blue $\textcolor{Aquamarine}{\bullet}$), CS4 (purple $\textcolor{Mulberry}{\bullet}$), potato starch PS (white o), and silica spheres SS (green $\textcolor{green}{\bullet}$).
        Guides to the eye are provided with dashed lines for select CS1 and CS3 data.} 
        \label{fig:stress-vs-velocity}
    \end{figure}

    Several aspects emerge from this figure.  
    Most importantly, for both setups, fracture can occur over a wide range of axial strains and normal stresses, above a threshold stress on the order of 0.1 MPa.
    When varying the different experimental parameters, we observe large scatter of the stress-strain data at fracture onset, which reflects the stochastic nature of the fracturing process.
    Despite this  scatter, it is observed that the stress at fracture increases with axial strain in both setups, with little to no dependence on volume fraction, as shown by the overlapping markers.
    Also, crack onset in the bulk setup (circles), where the shear-jammed material underneath the impactor exhibited yielding, is seen to generally require less stress than in the shallow setup (triangles).
    For some high volume fraction suspensions, the stress in shallow experiments is on the same order of magnitude as the bulk experiments.
    This crossover only occurs for low impact speed, where both shallow and bulk experiments find ductile yielding.
    As impact speed is not indicated in this figure, the dependence of stress on packing fraction is convoluted with its dependence on $v_{\mathrm{i}}$.
    The effect of impact speed on the minimum fracture stress and probability of fracture are discussed in detail in \textbf{Fig. \ref{fig:min_stress_at_fracture}} and \textbf{Fig. \ref{fig:stress-vs-velocity}}.
   
    To establish the fracture onset and its dependence on packing fraction and impact speed, we plot in \textbf{Fig. \ref{fig:min_stress_at_fracture}} the normal stress at which a suspension produced cracks with $>50$\% likelihood as a function of the particle volume fraction.
    To do so, we look at the lowest speed which yields fracture with $>50$\% likelihood and use the average fracture stress of all trials at this velocity to quantify a minimum stress for fracture at a given volume fraction.
    We see that this stress decreases with volume fraction, and that the velocity required to achieve $>50$\% likelihood of fracture also decreases as volume fraction increases.
    We observe this behavior in both setups, with greater decrease in stress in the shallow setup.
    Therefore, as previously observed from \textbf{Fig.\,\ref{fig:hist-and-scatter}}, stresses at fracture are larger in the shallow setup for a given volume fraction, and the impact speeds allowing for crack formation are also lower in the shallow container than in the bulk setup.

    \textit{Effect of Impact Speed} - To isolate the effect of impact speed on fracture onset, we pick a packing fraction and plot the corresponding subset of data as a function of $v_{\mathrm{i}}$ for each type of suspension.
    In \textbf{Fig. \ref{fig:stress-vs-velocity}} this is done by averaging CS1 data over a narrow range $\phi$ = 0.61-0.62 to obtain better statistics.
    This figure contains in addition data from experiments with low viscosity cornstarch (CS2, CS4), cornstarch with higher (CS2) or lower (CS3, CS4) surface tension, potato starch (PS) with the same solvent viscosity and surface tension as CS1 but roughly 3 times larger average particle diameter, and silica particles (SS) in the same solvent as CS1 but with smaller average diameter (see \textbf{Table \ref{tab:suspension-composition}}). The PS and SS suspensions have different volume fractions $\phi$ but are included in this comparison because their relative distance to volumetric jamming, given by $\phi/\phi_{\mathrm{J}}$, is similar (see Appendix B for a listing of $\phi_{\mathrm{J}}$ values).
    
    We first present in \textbf{Fig. \ref{fig:stress-vs-velocity}a} the probability of radial fracture as a function of impact speed.
    This probability rises once the impact speed exceeds a minimum value that depends on suspension type.
    Focusing on the 50\% midpoint of this rise, we find that data from shallow and bulk setup seem to track each other.
    Differences in particle size and/or shape do not produce a clear trend. 
    For example, the 50\% likelihood for observing fracture is achieved with similar $v_\text{i}$ in PS and SS, despite the average particle diameters differing by more than a factor 4. 
    Decreasing the viscosity or the surface tension of the suspending liquid (CS2 and CS3) shifts this midpoint to larger $v_\text{i}$ than CS1.
    The most pronounced shift from CS1 is an increase in required $v_\text{i}$, when both the surface tension $\gamma$ and the viscosity $\eta_0$ of the suspending liquid are reduced (CS4).

    In \textbf{Fig. \ref{fig:stress-vs-velocity}b} we plot the dependence of the normal stress at radial fracture $\sigma_\text{N}(z_\text{F})$, averaged over the same range of $\phi$, on the impact velocity $v_\text{i}$.
    As already seen in \textbf{Figs. \ref{fig:hist-and-scatter}} and \textbf{\ref{fig:min_stress_at_fracture}}, the typical normal stress at radial fracture for CS1 is  higher in the shallow setup, but for the lowest impact velocities yielding fracture, the stress drops down to the level observed in the bulk setup below 1 MPa (red triangles). 
    This decrease in $\sigma_\text{N}(z_\text{F})$ for the shallow setup correlates with the appearance of internal yielding, which we identify from the stress-strain curves (see \textbf{Fig. \ref{fig:yielding}a}).
    To indicate this, we plot fracture data  as stars when simultaneous internal yielding was not detected.
    A similar decrease in the fracture stress concomitant with the transition from internal strain stiffening to yielding behavior is also seen, at larger $v_\text{i}$, for silica spheres (SS) and potato starch (PS).
    We find that $\sigma_\text{N}(z_\text{F})$ varies with particle size, with suspensions using larger particles (PS) fracturing at lower average stresses than those using smaller particles.
    For all the cornstarch suspensions (CS) in the bulk setup, where we have the largest range of impact speeds and where the normal stresses at fracture can always be associated with internal yielding of the material under the impactor,  $\sigma_\text{N}(z_\text{F})$ follows the same increasing trend with $v_\text{i}$, roughly compatible with a power law with a value of 0.5 for the exponent.
    Interestingly, changing the surface tension and/or the viscosity of the solvent (CS2-4) has no noticeable effect on the magnitude of $\sigma_\text{N}(z_\text{F})$ and only changes the minimum required speed to reach that stress level.
    However, when the solvent viscosity was reduced (CS2, CS4) cracks could no longer be initiated in the shallow setup over the speed range tested.
    Few fracture events were observed in the silica suspension, and these occurred at relatively high $v_i$.
    
    \textbf{Figure \ref{fig:stress-vs-velocity}c} shows the hoop strain at fracture  $\varepsilon_\text{F}$, averaged over the same range of $\phi$, at fracture as a function of $v_\text{i}$.
    For CS1, we find that this tensile strain decreases roughly logarithmically with increasing impact speed, with both setups showing similar trends.
    The data for the other particle and solvent combinations, within scatter, are also compatible with a log-dependence on impact speed, although in some cases shifted to larger $v_\text{i}$.
    Increasing the particle size by switching to potato starch (PS), or increasing the surface tension while decreasing the solvent viscosity (CS2) is found to produce a more noticeable difference between the two setups, with the bulk setup requiring a larger hoop strain to generate fracture.
    Decreasing only the solvent viscosity and/or changing the surface tension (CS2, CS3, CS4) is seen to require, for given $v_\text{i}$, a larger hoop strain.

\section{\label{sec:Discussion}Discussion}

    Compared to indentation of ordinary elastic solids, the impact response of dense suspensions is more complex in that the application of local stress by the impactor not only deforms the material, but simultaneously drives a transition of state from fluid-like to solid-like.
    Current understanding has the transition from fluid into the SJ state under impact or extension occur at fixed $\phi$ in two stages as shear stress increases:
    first, a crossover from lubricated, effectively frictionless particle-particle contacts, to frictional particle interactions  leads to DST, resulting in a high-viscosity, yet, still fluid state \cite{wyart_discontinuous_2014,singh_constitutive_2018,morris_lubricated--frictional_2018}.
    Next, shear-jamming fronts convert fluid suspension into shear-jammed solid \cite{peters_direct_2016,han_high-speed_2016}.
    We see this first stage and the associated smooth deformation of the suspension surface around the impactor mainly in the bulk setup.
    In this configuration, the jamming fronts must travel over longer distances, so it takes more time to transform the material beneath the impactor into a shear-jammed solid., 
    Dense particulate media, such as the particle matrix in jammed suspensions, will necessarily dilate under shear \cite{jerome_unifying_2016}. 
    Upon entering the SJ regime, a suspension's surface will therefore transition from glossy to matte, since even slightly dilated particle configurations will lead to a roughened surface that produces diffuse scattering of light \cite{brown_shear_2014}.
    We observe this change of surface texture in our experiments as $z > z_{\mathrm{SJ}}$.

    Once the suspension has become jammed, further stress loading by uniaxial compression of the material directly underneath the impactor will lead to increasing dilation elsewhere.
    Therefore, as the impactor penetrates into the shear-jammed suspension, the hole widens and the material near the rim will come under tensile strain (hoop strain).
    Recent 3D time-resolved x-ray tomography by Dalbe \textit{et al.} of cohesive granular media  under impact, for which similar crack formation is observed,  quantifies this dilation and shows that the upper rim of the hole corresponds to the region of largest tensile strain \cite{dalbe-crack-2025}. 

    In our experiments, the onset of radial fracturing around the impact-induced hole can therefore be connected to the strength of the shear-jammed suspension material under tension, while tracking the normal stress as a function of penetration depth can show evidence of failure under compression.
    For the bulk setup and for low-speed impact in the shallow setup, these two types of failure are linked. 
    As \textbf{Fig. \ref{fig:yielding}b} shows, when the compressive strength is reached and the material shows signs of internal yielding, this yielding  generates sufficient dilation to also drive crack formation at the suspension surface.

    \textit{Mechanisms for Suspension Failure under Impact} - The observation that the cracks appear right at the suspension surface, radiating outward from the circumference of the hole punched by the impactor, suggests similarities with mode-1 fracture by blunt impact in other soft materials \cite{fakhouri_puncture_2015}, cohesive granular materials \cite{dalbe-crack-2025}, and amorphous solids more generally \cite{rattan_effect_2018}.
    In these situations the apparent stochastic nature of the cracking process results from its dependence on microscopic details of, and fluctuations in, material properties, in particular at the edge of the hole where the crack is initiated.
    
    Given that the rim of the hole is the region of largest tensile strain during penetration of the shear-jammed material \cite{dalbe-crack-2025}, it seems reasonable to associate the onset of radial fracture with a criterion for the maximum strain that adjacent particles can sustain before irreversibly separating.
    A strain threshold $\epsilon_{\mathrm{F}} \approx$ 0.10-0.12  was reported for radial fracture in impacted suspensions by Roch\'{e} \textit{et al.} \cite{roche_dynamic_2013} and for brittle failure under extension, by Bischoff White \textit{et al.} \cite{bischoff_white_extensional_2010} and also Smith \cite{smith_fracture_2015}.
    From \textbf{Fig. \ref{fig:stress-vs-velocity}c}, for the CS1 suspensions for which we have the most extensive data, it appears that the threshold hoop strain for fracture decreases from values around 0.1 to about 0.05 as the impact speed is increased  over two orders of magnitude. 
    This decrease appears to be approximately logarithmic  with impact speed (dashed lines).

    However, as suspension parameters such as the particle size, solvent viscosity or solvent surface tension are varied, the threshold strain for crack formation in \textbf{Fig. \ref{fig:stress-vs-velocity}c} exhibits behavior not captured by current theoretical models. 
    For partially wetted particles in granular packings, models for the cohesive stress between particles based on the existence of liquid bridges can predict failure strains around 0.1 \cite{dalbe-crack-2025}, but the limit of partial wetting is not applicable to the present situation of fully saturated suspensions.
    More appropriate models consider the cohesion between particles due to the capillary stress resulting from liquid menisci between particles at the suspension-air interface. 
    In dense particle packings the cohesive strength scales as $\gamma/d$, where $\gamma$ is the solvent's surface tension  \cite{roche_dynamic_2013,dufresne_dynamics_2006}.
    If capillary effects were significant for setting the threshold hoop strain, we would expect that changing $\gamma$ by $\pm$50\% between suspensions CS1-4 should produce a noticeable effect.
    \textbf{Figure 8c} shows that, for given impact speed, the hoop strain for fracture indeed changes, but in an unexpected way:
    as $\gamma$ is decreased and therefore the cohesive strength is reduced, the tensile strain required for fracture \textit{increases}. 
    This is seen in particular by comparing the data for CS2 and CS4.
    Effectively, therefore, the reduction in cohesive strength appears to make the dilating suspension less brittle and reduces the propensity for crack formation for given impact speed, requiring a larger tensile strain to fracture. 
    At the same time, changing $\gamma$ does not appear to affect the normal stress required to produce fracture (\textbf{Fig. \ref{fig:stress-vs-velocity}b}), suggesting that the strength of the shear-jammed material at the time of fracture is unrelated to surface tension.
     
    To account for strain rate dependence of the failure strain, Smith \cite{smith_fracture_2015} combined a rate-independent elastic strain corresponding to the brittle fracture limit around 0.1, with an empirical rate-dependent plastic strain tailored to capture the large plastic deformations observed in the highly ductile regime under slow extension.
    During extension, large plastic deformation leads to necking, but not necessarily to fracture, and can increase the failure strain dramatically, from $\approx$ 0.1 to values exceeding 0.5 \cite{smith_fracture_2015}.
    This differs from our observations for the hoop strain under impact, which, at the onset of crack formation, only reaches values up to 0.15 even at the lowest impact speeds (\textbf{Fig. \ref{fig:stress-vs-velocity}c}), and thus involves considerably less contribution from plastic deformation.
    In comparison to the suspensions used by Smith \cite{smith_fracture_2015}, our lower critical hoop strains indicate that our suspensions are more brittle and closer to the limit of failure.
    Also, the mild, approximately logarithmic dependence of the hoop strain on impact speed we see in the CS1 data (and shifted to larger speeds for CS2-4) seems to indicate that the shear-jammed suspension becomes more brittle with increasing $v_{\mathrm{i}}$.

    In order to capture the rate dependence of the response to impact, a microstructure-based approach has been to treat the jammed particulate phase of the suspension as a porous medium and consider the pore pressure. 
    From Darcy's Law, the pressure required to force a liquid of viscosity $\eta_0$ at a speed $v$ through a porous medium over a depth $L$ and permeability $\kappa$, scales as $\eta_0 vL/\kappa$.
    For a bed of randomly packed monodisperse spheres of diameter $d$, $\kappa$ is typically represented by the Kozeny-Carman relation $\kappa = (1 - \phi^3) d^2 / 180 \phi^2$.
    Accounting furthermore for the coupling of pore pressure and dilation, as appropriate for failure of a densely packed, or jammed particle bed, Jerome et al. \cite{jerome_unifying_2016} arrive at a pore pressure of the form  $\alpha\eta_0 v_\text{i} \Delta \phi L / \kappa$, where $\alpha$ is a constant of order unity (in Ref. \cite{jerome_unifying_2016}, $\alpha = $4), and $\Delta\phi = \phi - \phi_c > 0$ indicates the extent to which the initial suspension packing fraction, $\phi$, exceeds the value $\phi_c$ that delineates a response to applied stress characterized by dilation from one characterized by compaction of the particle packing.
    Accounting for the frictional interactions in the granular matrix under uniaxial compression leading to dilation, the resistance to impact, i.e. the stiffness of the material, the pore pressure is then ten times larger that that given by Darcy's law \cite{jerome_unifying_2016}. 
    Taking $\phi_c \approx 0.58$ as in Refs. \cite{jerome_unifying_2016, dufresne_dynamics_2006}, we find $\Delta\phi \approx 0.01-0.03$ for our experiments.
    Assuming that the fluid moves at impact speed $v_\text{i}$, and taking $L = 2a$ as the characteristic distance over which the impact stress is spread out, this predicts material stiffnesses or effective moduli that are strongly speed dependent, ranging from $\approx$1kPa at 0.1 mm/s to $\approx$ 200 kPa at 10 mm/s.

\begin{figure}
        \centering
        \includegraphics[width=0.9\linewidth]{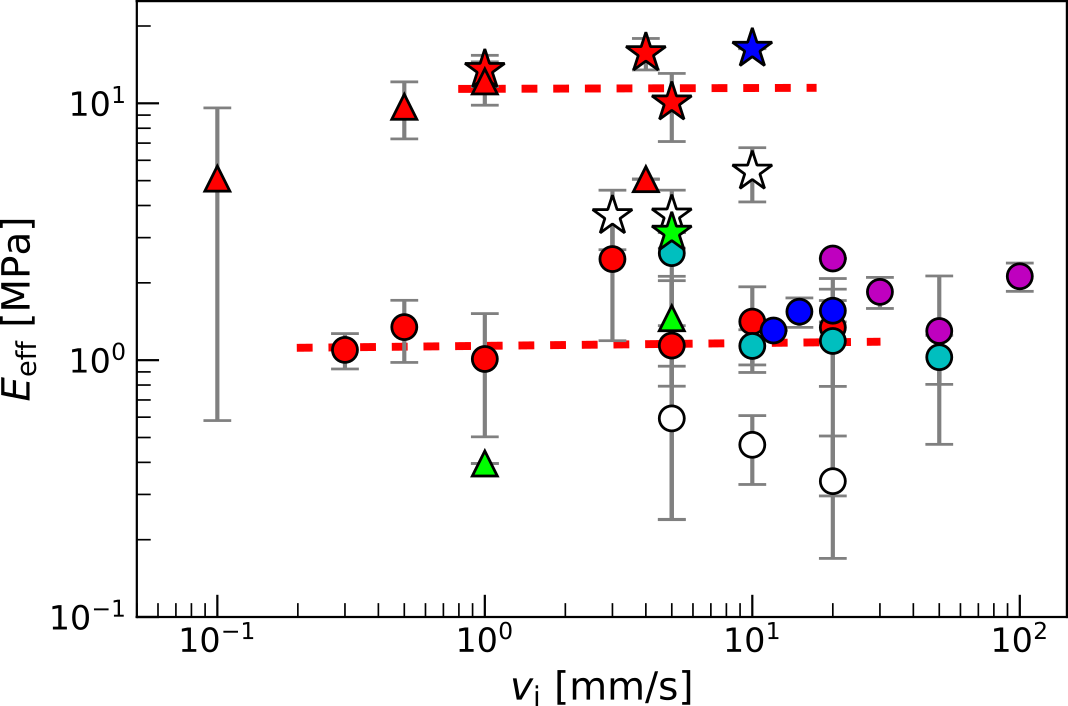}
        \caption{Effective modulus at onset of fracture $E_{\mathrm{eff}}$ as a function of impact speed $v_\text{i}$. The color and shape of the markers are the same as in \textbf{Fig. \ref{fig:stress-vs-velocity}}.
        Guides to the eye are provided with dashed lines for select CS1 data.}
        \label{fig:modulus}
    \end{figure}

    In the experiments, we can obtain an effective modulus from the slope of stress-strain curves.
    Just before the onset of radial cracks it is given by $E_{\mathrm{eff}}=\left[H\mathrm{d}\sigma_{\mathrm{N}}/\mathrm{d}z\right]_{z=z_{\mathrm{F}}}$. 
    This is plotted in \textbf{Fig. \ref{fig:modulus}}, where $E_{\mathrm{eff}}$ clusters around two values about an order of magnitude apart: a value of $\approx$1 MPa for shear-jammed material that shows yielding and ductile behavior, and a value of $\approx$10 MPa for material that exhibits strain-stiffening.
    These values appear, on average, remarkably independent of impact speed, except  in the shallow setup for $v_{\textrm{i}} <$ 1 mm/s, where we observe the speed-dependent crossover from strain-stiffening (stars) to internal yielding (triangles) behavior. 
    The values are in the same range $10^5 - 10^7$ Pa as those reported by Roch\'{e} \textit{et al.} \cite{roche_dynamic_2013} and Maharjan \textit{et al.} \cite{maharjan_constitutive_2018} for shear-jammed cornstarch suspensions under impact.
    However, our data in \textbf{Fig. \ref{fig:modulus}} differ from those in Maharjan \textit{et al.}, who reported a speed dependence $E_{\mathrm{eff}} \propto v_{\textrm{i}}$ of the modulus, with a confinement-dependent saturation at speeds $v_{\textrm{i}} >$ 100 mm/s.
    
    An increasing, linear dependence of the effective modulus $E_{\mathrm{eff}}$ on impact speed $v_{\mathrm{i}}$, predicted by models based on Darcy's Law therefore is not compatible with our experimental data in \textbf{Fig. \ref{fig:stress-vs-velocity}}. 
    Furthermore, given the $d^{-2}$ dependence of the pore pressure on particle diameter, such models would predict the larger potato starch particles ($d = 42$ $\mu$m) to exhibit a significantly lower modulus than cornstarch ($d = 14$ $\mu$m). 
    However, the variation we observe for $E_{\mathrm{eff}}$ is within a factor of 2.
    In addition, switching to a suspending liquid with five times smaller viscosity (CS2) did not lead to a noticeably lower $E_{\mathrm{eff}}$.  
    
    As a result, currently available approaches for granular media or dense suspensions that are based on capillary cohesion or pore pressure appear to be unable to predict correctly the trends $\sigma_\text{N}(z_\text{F})$, $\varepsilon_\text{F}$, and $E_{\mathrm{eff}}$  we observe experimentally.

    \textit{The Role of Confinement} - While impact using the bulk setup led to yielding of the material underneath the impactor, it was possible in the shallow setup to track the crossover from yielding behavior to strain-stiffening, and thus more brittle impact response, by increasing $v_{\mathrm{i}}$ (see \textbf{Figs. \ref{fig:experimental-setup}e} and \textbf{\ref{fig:yielding}a}).  
    Since this change in behavior correlates with the change in stiffness right before failure $E_{\mathrm{eff}}$,  we associate the two modulus values for CS1 in \textbf{Fig. \ref{fig:modulus}}, $E_{\mathrm{eff}}\approx$1 MPa and $E_{\mathrm{eff}}\approx$10 MPa, with different impact-induced structures of the network of frictional contact forces in the shear-jammed suspension.
    In the shallow setup, the smaller suspension volume allows for more uniform jamming throughout the suspension, and reflection of the shear-jamming fronts from nearby walls can generate a more isotropic force chain fabric.
    In addition, applied stress is more easily shunted toward the walls, which increases the ultimate strength of the shear-jammed material, which therefore strain stiffens.

    That different boundary conditions can lead to different stress-activated force chain networks (fabrics), and thus material stiffnesses, was also noted by Petel and Ouellet \cite{petel_dynamic_2017} for high-speed impact.
    It furthermore could explain why not only the stiffness but also the failure and fracture strengths are lower when wall confinement is absent, such as under extension, where these strengths are in the range of $10^4 - 10^5$ Pa  \cite{smith_dilatancy_2010, smith_fracture_2015, bischoff_white_extensional_2010}.
    In \textbf{Fig. \ref{fig:min_stress_at_fracture}}, at a given $\phi$, the minimum stress $\sigma_\text{N}(z_\text{F,50\%})$ for fracture onset shows differences (up to a factor 3) between the two setups, especially at low packing fraction. 
    Since in both cases this onset stress is correlated with internal yielding of the very same material, we similarly associate the higher strength for the shallow setup with impact establishing a more interconnected, isotropic force chain network. 
    We can expect this difference to become less significant for larger $\phi$, where the material gets jammed more uniformly over a larger extent, which is consistent with the data.

    The same boundary and impact conditions that produce a shear-jammed solid of a particular stiffness also determine how the surface of such solid will deform under impact.
    Conversely, from its surface deformation, we can obtain information about the material properties of the shear-jammed solid.
    We can get estimates for the material properties thanks to the aforementioned COMSOL simulation and the data in \textbf{Fig. \ref{fig:laser-fig}c,d}, when the shear-jammed suspension undergoes puncture.
    The puncture onset is defined as the depth at which the suspension surface profile local to the impactor ($r = 6$ mm) diverges from the motion of the impactor, corresponding to impactor depths of $z\geq 2.2$~mm.
    Comparing the simulation data with the experimentally observed displacement of the suspension at different distances from the impactor, we find that good agreement is obtained for ratios of bulk modulus, $B$, to shear modulus, $G$, of $100{:}1$ or larger, e.g., $B/G = 1000$ to obtain the green solid lines in \textbf{Fig. \ref{fig:laser-fig}}.
    This ratio is related to the Poisson's ratio of the material through $B/G = \tfrac{2}{3}(1 + \nu)/(1 - 2\nu)$.
    Simulations therefore indicate that surface deformation of the material at this time of the impact is indicative of Poisson's ratios $\nu \geq 0.495$, corresponding to the nearly incompressible limit (due to the water content of the suspensions).
    From \textbf{Fig.~\ref{fig:modulus}}, the effective Young’s modulus of the shear-jammed suspension in the bulk setup is $E_{\mathrm{eff}} \approx 1$~MPa for CS1.
    Knowing that $B = E/[3(1 - 2\nu)]$ and $G = E/[2(1 + \nu)]$, we get $B \approx 30$~MPa and $G\approx 0.3$~MPa, taking $B/G = 100$.
    The solid green lines in \textbf{Fig. \ref{fig:laser-fig}c,d} show the best fit between simulations and experiments during the portion of the impact in which the suspension exhibits puncture.
    This analysis therefore indicates that the shear-jammed suspension undergoing puncture behaves as a nearly incompressible solid.

\begin{figure}[h]
        \centering
        \includegraphics[width=\linewidth]{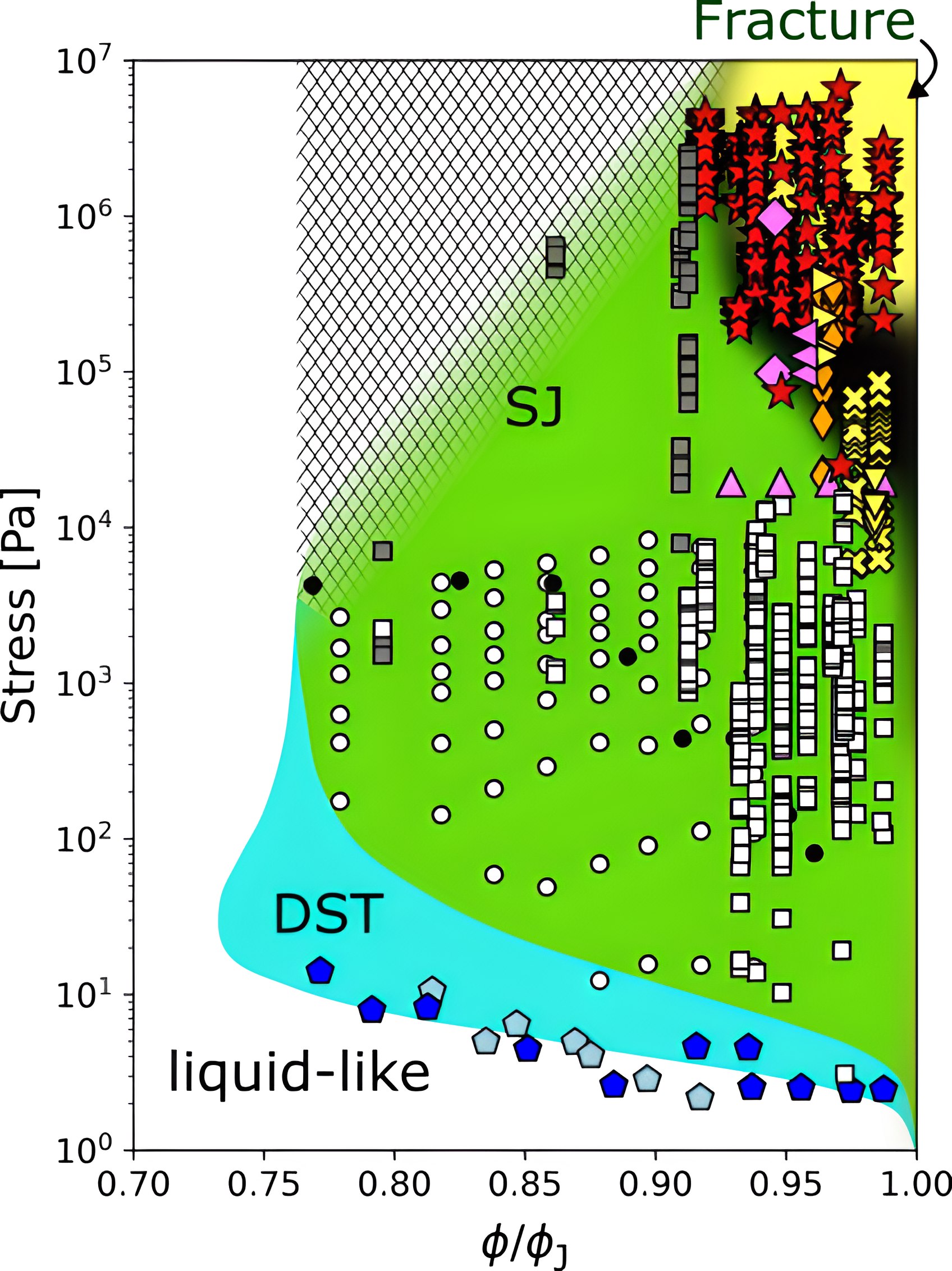}
        \caption{State diagram for dense suspensions. Shading indicates suspension state: white, liquid-like; blue, DST; green, SJ;  yellow/black, fracture. Data shown from this work include the normal stress from \textbf{Fig. \ref{fig:laser-fig}-\ref{fig:min_stress_at_fracture}} at the onset of radial fracture under impact (red stars), the normal stress at $z_\text{SJ}$ (white squares), and the maximum normal stress in non-fracturing trials (gray squares). 
        Data for fracture due to gas invasion is from Ozturk \textit{et al.} \cite{ozturk_flow--fracture_2020} (wide pink diamonds), Singh \textit{et al.} \cite{singh_viscous_2023} (pink left-facing triangles) and Lilin \textit{et al.} \cite{lilin_fracture_2024} (pink upward-facing triangles).
        Data for fracture in ballistic experiments is from Petel \textit{et al.} \cite{petel_dynamic_2017} (orange diamonds).
        Data for failure under extension is from Bischoff White \textit{et al.} \cite{bischoff_white_extensional_2010} (yellow downward-facing triangles), Smith \cite{smith_fracture_2015} (yellow x's), and Chen \textit{et al.} \cite{chen_leveraging_2023} (yellow right-facing triangles).
        The border of the blue DST region is drawn according to the model by Wyart and Cates \cite{wyart_discontinuous_2014}.
        Data delineating the onset of DST and SJ are from Han \textit{et al.} \cite{han_shear_2018, han_stress_2019} (light blue pentagons, white circles) and Peters \textit{et al.} \cite{peters_direct_2016} (dark blue pentagons, black circles).
        The cross-hatched region of the plot is an area yet to be probed.}
        \label{fig:stress-vs-phi}
    \end{figure}

    \textit{State Diagram} - The state diagram for dense suspensions delineates liquid-like (Newtonian or shear thinning) behavior from shear thickening and shear jamming states, as a function of volume fraction and applied stress \cite{wyart_discontinuous_2014, singh_constitutive_2018,han_stress_2019,peters_direct_2016}.  
    In \textbf{Fig. \ref{fig:stress-vs-phi}}, adapted from Han \textit{et al.} \cite{han_stress_2019}, the boundaries for discontinuous shear thickening (DST, blue) and shear jamming (SJ, green) are based on the work by Wyart and Cates \cite{wyart_discontinuous_2014}.
    The volume fraction $\phi$ in this plot has been normalized by the volume fraction $\phi_\text{J}$ for volumetric jamming  in the quiescent state without applied stress, which for suspensions corresponds to the jamming of fully lubricated and thus frictionless particles \cite{wyart_discontinuous_2014}.
    For the transitions from fluid-like behavior into the DST or SJ regime, the relevant stress variable is the shear stress.  
    In the solid-like SJ regime, we use the normal stress $\sigma_{\mathrm{N}}(z_{\mathrm{F}})$ as a proxy for the shear stress and populate this diagram with our complete set of data from both setups.
    
    We find that the data for impact-induced radial fracture (red stars) fall into a distinct region at high normalized packing fraction and high stress.
    Within this region, the fracturing process is stochastic.
    The region shaded yellow indicates the regime in which we observed fracture with $>$50\% likelihood even at the lowest impact speed.
    Its lower boundary is determined from data in \textbf{Fig. \ref{fig:min_stress_at_fracture}}.
    The darker band around the edge of the yellow fracture regime represents a transition regime, where fracture was observed with lower probability.
    Fracture data from experiments with low impact speed or from less confined conditions (bulk setup) populate the low-stress end near that boundary.
    Indeed, if it occurred, the internal yielding of the material under the impactor produced the minimum required dilation and therefore a lower stress than that produced by the same material undergoing strain stiffening.    
    Therefore, the boundary of this fracture regime also delineates the lower limit for shear-jammed suspensions to yield under compression. 
    At the low-$\phi$ side of the fracture regime, corresponding to $\phi$ = 0.59 in our experiments, the boundary is formed by the vertical line of dark gray squares. 
    These gray squares indicate the maximum stress value for trials where fracture was not observed.
    
    The white squares in \textbf{Fig. \ref{fig:stress-vs-phi}} mark where we first detect shear jamming in the traces of normal stress vs. axial strain (see \textbf{Figs. \ref{fig:experimental-setup}d,e)}. 
    These data points populate the green SJ regime.
    Given the complex stress field generated by impact, the method employed here cannot accurately delineate the boundary between the DST and SJ regimes.
    For that, more controlled experiments are required \cite{han_shear_2018,han_stress_2019}.
    We can, however, use the increase in normal stress to determine when the suspension has transformed into its solid-like SJ state, and our data overlap with similar data from Refs. \cite{han_stress_2019,peters_direct_2016} (white and black circles) that extend to lower packing fraction.
    For completeness, we also plot experimental data from Refs.\cite{han_stress_2019,peters_direct_2016} delineating the transition into DST from liquid-like behavior or mild, continuous shear thickening (dark and light blue pentagons).
    
    In this state diagram, our data for the stress associated with impact-induced radial cracking can be compared to data obtained by other means of inducing fracture or failure in dynamically jammed suspensions.
    For similar impact experiments, Roch\'{e} \textit{et al.} \cite{roche_dynamic_2013} reported $\phi/\phi_\mathrm{J} > 0.9$, and Allen \textit{et al.} \cite{allen_system-spanning_2018} and Maharjan \textit{et al.} \cite{maharjan_constitutive_2018} reported $\sigma_{\mathrm{N}}(z_{\mathrm{F}}) > 4 \times 10^6$ Pa for fracture, and therefore their data fall in the same region in this diagram.
    Interestingly, high-velocity data by Petel \textit{et al.} for impact in cornstarch or silica suspensions at $v_\mathrm{i} = 200 - 700$ m/s show stress levels for yielding that are comparable and can even be somewhat lower  (orange diamonds; \cite{petel_dynamic_2017}). 
    This supports the notion that the modulus as well as the strength of shear-jammed materials tend to saturate, possibly already at the low impact speeds in our experiments.
    Data by Ozturk \textit{et al.} \cite{ozturk_flow--fracture_2020} for fracture via pressurized gas invasion in Hele-Shaw cells (wide pink diamonds) similarly overlap with our results.
    As the packing fraction $\phi$ approaches $\phi_\text{J}$, the fracture regime (yellow) extends to lower stress levels (see \textbf{Fig. \ref{fig:min_stress_at_fracture}}), and associated with that the likelihood increases for fracture events at stresses significantly lower than $\sigma_\text{N}(z_\text{F,50\%})$, as seen from the data in \textbf{Fig. \ref{fig:hist-and-scatter}} and shown here by an extended black region.
    Low fracture stresses at large $\phi$ can also be found when columns of dense suspensions are subjected to extension. 
    This is shown in \textbf{Fig. \ref{fig:stress-vs-phi}} by data from Bischoff White \textit{et al.} \cite{bischoff_white_extensional_2010}, Smith \cite{smith_fracture_2015}, and Chen \textit{et al.} \cite{chen_leveraging_2023}.
    As noted earlier, we speculate that this may be due to the absence of nearby walls and thus less confinement, which produces a shear-jammed state that is both softer and less strong.
    This may also explain the lower stress level for fracture via gas invasion found in bulk volumes of suspension by Lilin \textit{et al.} (upward-facing pink triangles; \cite{lilin_fracture_2024}), although these experiments observed fracture to extend to lower $\phi$ than any other reports in the literature.
    
    The region above the SJ regime and to the left of the fracture regime is cross-hatched in this figure, as the behavior of dense suspensions under these conditions has not yet been explored systematically.
    Dense suspensions in this region of the state diagram  likely still fail at high stresses, but the low-$\phi$ limit of our fracture regime implies that this failure may not include radial cracks.

\section{\label{sec:Conclusions}Conclusions}

    Under impact, dense, initially liquid-like suspensions can be forced into a solid-like shear-jammed state. 
    Once in this state, further straining by the impactor will dilate the shear-jammed material to the point of failure. 
    Focusing on signatures of this failure, we observe the formation of radial cracks starting from the rim of the hole punched into the suspension by the impactor, similar to what occurs in brittle solids.
    Using suspensions of different starch particles as well as polydisperse silica spheres we observe the onset of cracking to be a stochastic process, occurring across a wide range of normal (axial) stresses and axial strains produced by the impactor (\textbf{Fig. \ref{fig:hist-and-scatter}}), and we extract minimum criteria that need to be satisfied for likely crack formation. 
    Our data show that a shear-jammed suspension can develop cracks for packing fractions $\phi >$ 0.59 and once it has dilated sufficiently for the tensile hoop strain around the hole to exceed a threshold value $\varepsilon_{\text{F}}$. 

    In the absence of predictive models for the strength of dynamically shear-jammed suspensions, our results provide baseline data that extend the findings in Refs. \cite{roche_dynamic_2013,allen_system-spanning_2018, maharjan_constitutive_2018}.
    In particular, our experiments demonstrate how speed-controlled impact can be used to establish the conditions and control the likelihood for radial cracks to occur (\textbf{Fig. \ref{fig:stress-vs-velocity}}). 
    The minimum impact speed for 50\% likelihood of radial fracture, while not particularly sensitive to particle diameter $d$, is found to increase with decreased solvent viscosity $\eta_0$ as well as when the solvent's surface tension $\gamma$ is reduced. 
    Under these conditions the hoop strain at fracture also increases, indicating that the jammed material has become less brittle.
    We find that the normal stress at fracture, for given particle type and packing fraction, is affected mainly by whether or not the suspension material underneath the impactor undergoes ductile yielding while being compressed.
    We observe such yielding in cases where the shear-jammed material has a low effective modulus $E_{\mathrm{eff}}$, which we associate with conditions that produce a less interconnected and thus weaker network of frictional contact forces among particles, such as slow impact speed and/or  less lateral confinement of the volume of suspension underneath the impactor.

    The onset of crack formation can be used to delineate a new fracture regime  in the state diagram for dense suspensions. 
    The compilation of data in \textbf{Fig. \ref{fig:stress-vs-phi}}, obtained for different suspension types and different experimental conditions,  significantly extends the current state diagram to higher stress levels and outlines where the shear-jammed solid reaches its load-bearing capacity and starts to fail either under extension or compression, with the remarkable ability to fracture similar to a brittle solid.
    Finally, the minimum packing fraction of 0.59 for the fracture regime implies that shear-jammed suspensions with slightly lower $\phi$ should fail in a different manner, and this suggests the possibility that additional failure regimes atop the SJ region could be identified. 

\section{Acknowledgments}

    We thank Samantha Livermore and Hojin Kim for many helpful discussions and Jochem Meijer for assistance with the COMSOL simulations.
    This work was supported by the University of Chicago Materials Research Science and Engineering Center (MRSEC), which is funded by the National Science Foundation under award number DMR-2011854.
    M. S. acknowledges a MRSEC Graduate Student Fellowship and A. P. support from the Army Research Office under grants W911NF-21-2-0146 and from the Kadanoff-Rice Postdoctoral Fellowship. H.M.J. acknowledges support from the Army Research Office under grant W911NF-25-2-0178.

\section{\label{sec:Appendix A}Appendix A: Calculation of suspension volume fraction}

    Starch suspensions were prepared with particle weight fraction $0.50 < \phi_\text{p} < 0.56$, which was then converted to particle volume fraction $0.51 < \phi < 0.63$ by considering the moisture content $\beta$ and porosity $\xi$ of the suspended particles, through the relationship \cite{han_stress_2019}
    \begin{equation}
    \large
        \phi = \frac{1}{1 - \beta} \frac{(1 - \xi) \left(\frac{\phi_\text{p}}{\rho_\text{p}} \right)}{(1 - \xi) \left( \frac{\phi_\text{p}}{\rho_\text{p}} \right) + \frac{1 - \phi_\text{p}}{\rho_\text{l}} + \xi \left( \frac{\phi_\text{p}}{\rho_\text{w}} \right)}.
    \end{equation}
    
    The density of the liquid solvent is given by $\rho_\text{l}$, and $\rho_\text{w} = 1000$ kg/m$^3$ is the density of water. 
    Values used for each particle and solvent are provided in \textbf{Tables\,\ref{tab:SI_particles},\ref{tab:SI_solvents}}.
    
\begin{table}[h]
\centering
\renewcommand{\arraystretch}{1.5}
\begin{tabular}{l c c c}
\hline
Particle & Density $\rho_\text{p}$ [kg/m$^3$] & Moisture Content $\beta$ & Porosity $\xi$ \\
\hline \hline
Cornstarch & 1400 & 0.14 & 0.31 \\
Potato starch & 1500 & 0.15 & 0.31 \\
Silica & 1100 & 0.0 & 0.0 \\
\hline
\end{tabular}
\caption{Particles used in the experiments}
\label{tab:SI_particles}
\end{table}

\begin{table}[h]
\centering
\renewcommand{\arraystretch}{1.5}
\begin{tabular}{l c}
\hline
Solvent & Density $\rho_\text{l}$ [kg/m$^3$] \\
\hline \hline
1:1 glycerol:DI water & 1130 \\
37:63 CsCl:DI water & 1400 \\
\makecell[l]{1:1 glycerol:DI water \\ with Dawn dish soap} & 1130 \\
\makecell[l]{DI water with Dawn \\ dish soap} & 1000 \\
\hline
\end{tabular}
\caption{Solvents used in the experiments}
\label{tab:SI_solvents}    
\end{table}

\section{\label{sec:Appendix B}Appendix B: Volumetric jamming packing fractions}

    For the suspensions used in this work, $\phi_\text{J}$ was determined experimentally by adding particles until the mixture no longer flowed in its undisturbed state.
    For data taken from the literature, $\phi_\text{J}$ was either provided or estimated by referencing similar papers or making reasonable assumptions about the size and type of suspended particles and solvent, especially using Ref. \cite{han_measuring_2017} for suspensions containing cornstarch.

\begin{table}[h]
\centering
\renewcommand{\arraystretch}{1.5}
\begin{tabular}{l c}
\hline
Suspension & $\phi_\text{J}$ \\
\hline \hline
CS1 & 0.65 \\
CS2 & 0.64 \\
CS3 & 0.65 \\
CS4 & 0.65 \\
PS & 0.62 \\
SS & 0.66 \\
\makecell[l]{Han \textit{et al.} \\ \cite{han_shear_2018, han_stress_2019}} & 0.65 \\
\makecell[l]{Peters \textit{et al.} \\ \cite{peters_direct_2016}} & 0.62 \\
\makecell[l]{Petel \textit{et al.} (CS) \\ \cite{petel_dynamic_2017}} & 0.56 \\
\makecell[l]{Petel \textit{et al.} (SS) \\ \cite{petel_dynamic_2017}} & 0.63 \\
\makecell[l]{Chen \textit{et al.} \\ \cite{chen_leveraging_2023}} & 0.58 \\
\makecell[l]{Smith \textit{et al.} \\ \cite{smith_dilatancy_2010, smith_fracture_2015}} & 0.64 \\
\makecell[l]{Ozturk \textit{et al.} \\ \cite{ozturk_flow--fracture_2020}} & 0.56 \\
\makecell[l]{Singh \textit{et al.} \\ \cite{singh_viscous_2023}} & 0.58 \\
\makecell[l]{Lilin \textit{et al.} \\ \cite{lilin_fracture_2024}} & 0.64 \\
\makecell[l]{Bischoff White \textit{et al.} \\ \cite{bischoff_white_extensional_2010}} & 0.58 \\
\hline
\end{tabular}
\caption{Jamming volume fractions $\phi_\text{J}$}
\label{tab:phi_J}
\end{table}

\bibliography{library_clean}

\end{document}